\documentclass{article}
\usepackage{ijcai26}

\usepackage{times}
\usepackage{soul}
\usepackage{url}
\usepackage[hidelinks]{hyperref}
\usepackage[utf8]{inputenc}
\usepackage[small]{caption}
\usepackage{graphicx}
\usepackage{amsmath}
\usepackage{amsthm}
\usepackage{booktabs}
\usepackage{algorithm}
\usepackage{algorithmic}
\usepackage[switch]{lineno}
\usepackage{amssymb}
\usepackage{multirow}
\usepackage{makecell}
\usepackage{amsthm}              
\newtheorem{proposition}{Proposition}  

\newtheorem{theorem}{Theorem}

\title{Bridging the Gap Between Estimated and True Regret: Towards Reliable Regret Estimation in Deep Learning-based Mechanism Design}

\author{
Shuyuan You$^1$
\and
Zhiqiang Zhuang$^2$\and
Kewen Wang$^{1}$\And
Zhe Wang$^{1}$\\
\affiliations
$^1$Griffith University\\
$^2$Qiannan Normal University for Nationalities\\
\emails
shuyuan.you@griffithuni.edu.au,
Zoldlady@gmail.com,
k.wang@griffith.edu.au,
zhe.wang@griffith.edu.au
}

\begin{document}

\maketitle

\begin{abstract}

Recent advances, such as RegretNet, ALGnet, RegretFormer and CITransNet, use deep learning to approximate optimal multi-item auctions by relaxing incentive compatibility (IC) and measuring its violation via ex-post regret. However, the true accuracy of these regret estimates remains unclear. Computing exact regret is computationally intractable, and current models rely on gradient-based optimizers whose outcomes depend heavily on hyperparameter choices. Through extensive experiments, we reveal that existing methods systematically underestimate actual regret (In some models, the true regret is several hundred times larger than the reported regret), leading to overstated claims of IC and revenue. To address this issue, we derive a lower bound on regret and introduce an efficient item-wise regret approximation. Building on this, we propose a guided refinement procedure that substantially improves regret estimation accuracy while reducing computational cost. 
Our method provides a more reliable foundation for evaluating incentive compatibility in deep learning–based auction mechanisms and highlights the need to reassess prior performance claims in this area.
\end{abstract}

\section{Introduction}



Optimal auction design is a central problem in algorithmic game theory with substantial practical relevance, including applications in spectrum allocation \cite{cramton2013spectrum} and online advertising \cite{liu2021neural}. While Myerson \cite{myerson1981optimal} characterised the optimal auction for the single-item setting, the problem becomes substantially more complex for multiple distinct items. Despite extensive research \cite{DBLP:conf/sigecom/AlaeiFHHM12,babaioff2020simple,cai2017simple,hart2017approximate,pavlov2011optimal,wang2014optimal,yao2017dominant}, the exact optimal mechanism remains unknown even for as few as two items and two bidders.

Given this theoretical intractability, recent work has turned to deep learning–based approaches. A representative example is RegretNet \cite{dutting2019optimal}, which parameterises auction mechanisms using neural networks. Because training relies on back-propagation and gradient descent, enforcing incentive compatibility (IC) directly is challenging. Instead, these models relax the IC constraint and quantify violations via ex-post regret, defined as the maximum utility gain a bidder can obtain by deviating from truthful reporting. Those learning-based approaches achieved higher revenue than traditional methods\cite{palfrey1983bundling}, while ensuring that regret remained at a very low level.

However, a critical yet underexplored question remains: do these models accurately estimate incentive-compatibility violations, as measured by regret? Accurate regret estimation is fundamental to deep learning–based mechanism design, as it underpins the credibility of reported revenue performance. When the regret optimisation procedure fails to identify the misreport that leads to the largest utility gain, the resulting regret estimates can be substantially lower than the actual regret. A mechanism may then appear to achieve high revenue, even though such conclusions are theoretically unsound and practically unreliable.

From a theoretical perspective, computing a bidder’s optimal ex-post regret requires searching over the entire combinatorial valuation space while holding other bidders’ reports fixed. This task suffers from the curse of dimensionality, growing exponentially with the number of items and rendering exact computation intractable. As a result, existing models rely on gradient-based optimisation to approximate the optimal misreport. However, such approximations are highly sensitive to hyperparameters—including the number of random initialisations, gradient steps, and learning rate. This sensitivity gives rise to two fundamental limitations:

\emph{Estimation Inaccuracy}:
Existing models lack a rigorous assessment of whether gradient-based approximations converge to values close to the true regret. Owing to the highly non-convex loss landscape, regret computation is extremely sensitive to hyperparameter choices. Suboptimal configurations can therefore lead to systematic underestimation of regret, creating a false impression of incentive compatibility. Although recent work such as RegretFormer \cite{ivanov2022optimal} introduces diagnostic consistency checks, these criteria are necessary but not sufficient for recovering the true regret.

\emph{Computational Inefficiency}:
Our empirical analysis reveals that increasing the number of random initialisations and gradient steps can improve estimation accuracy, but at the cost of prohibitive computational overhead. In large-scale settings, obtaining a converged regret estimate may require hundreds of hours—far exceeding the time needed to train the mechanism itself. This efficiency bottleneck highlights the urgent need for more efficient regret evaluation methods.

To address the critical challenge of regret underestimation, we develop a theoretically grounded and computationally efficient evaluation framework. We first derive a mathematical lower bound on the optimal regret and introduce \emph{Item-wise Regret} as a novel proxy metric. This approach decomposes the intractable joint optimisation problem into independent component-wise subproblems, reducing the computational complexity from exponential, $O(Q^m)$, to linear, $O(m)$.

Importantly, we go beyond simple discretisation by introducing an Item-wise Guided Gradient Refinement strategy. Unlike conventional approaches that rely on random initialisations, our method uses discrete item-wise solutions to construct a structured set of initial points for gradient-based optimisation. This design bridges discrete grid search and continuous optimisation, substantially improving the optimiser’s ability to converge to solutions closer to the global optimum. Our framework offers two key advantages:

\textbf{Improved Regret Estimation}:
Our method uses item-wise solutions to guide the optimiser toward regions corresponding to high regret. This mitigates common failure modes of random initialisation, where the optimiser becomes trapped in poor local optima, and enables reliable regret estimation even in complex, high-dimensional auction settings.

\textbf{Orders-of-Magnitude Efficiency Gains}:
The structured initialisation substantially reduces sample complexity. In large-scale settings, where exhaustive or random search methods may require thousands of random restarts and hundreds of hours of computation, our framework achieves superior convergence within a few hours using only dozens of strategically chosen initialisations. This reduction in computational overhead makes practical and scalable regret evaluation of deep auction mechanisms feasible.

\textbf{Related work}

RegretNet \cite{dutting2019optimal} was the first model to utilize neural networks for approximating near-optimal auctions. Several follow-up works have extended and refined these models.ALGnet~\cite{rahme2021auction} reformulates the learning process as a two-player game between the auctioneer and a misreporter. Feng et al.\ \cite{feng2018deep} addressed private budget constraints, while Kuo et al.\ \cite{kuo2020proportionnet} introduced fairness. Peri et al.\ \cite{peri2021preferencenet} incorporated human preferences into auction design. Rahme et al.\ \cite{rahme2021permutation} developed permutation-equivariant mechanisms, and Curry et al.\ \cite{curry2022differentiable} revisited affine maximizer auctions. Other notable contributions include Shen et al.\ \cite{shen2019automated}, Dütting et al.\ \cite{dutting2019optimal}, and Duan et al.\ \cite{duan2022context}, each addressing specific aspects of strategyproofness, accuracy, and contextual information integration in auction design. You et al. \cite{you2023improving} proposed two methods for improving ALGnet with discrete distribution sampling and a payment network fine-grained to achieve state-of-the-art experimental results. Wang et al. \cite{wang2024gemnet} extended the menu-based approach of the single-bidder RochetNet to the multi-bidder setting. Huo et al. \cite{huo2025learning} proposed a new auction mechanism to deal with the correlation of values.

\section{Preliminaries}
We consider a scenario where $n$ bidders are present, represented by the set $N = \{1, \ldots, n\}$, and $m$ items are available for auction, represented by the set $M = \{1, \ldots, m\}$. Each bidder $i$ has a valuation function $v_i: 2^M \rightarrow \mathbb{R}_{\geq 0}$.We focus on \textit{additive} valuation functions, that is, a bidder $i$'s valuation for the set $S\subseteq M$ of items is the sum of the valuations for each item in $S$.
Formally, $v_{i}(S)=\sum_{j\in S}{v_{i}(j)}$ for all $S\subseteq M$, where $v_i$ is independently drawn from a distribution $F_i$ over possible valuation functions $V_i$. We denote the set of all possible valuation profiles as $V = \prod_{i=1}^n V_i$, and $v = (v_1, \ldots, v_n) \in V$ is the valuation profile. 

Although the auctioneer knows the distributions $F = (F_1, \ldots, F_n)$, the actual valuations $v$ of the bidders are unknown. Bidders report their bids, which may not be truthful, and an auction mechanism determines the allocation of items to the bidders and the corresponding payments. An auction mechanism $(g, p)$ is a combination of allocation rules $g_i : V \rightarrow 2^M$ and payment rules $p_i : V \rightarrow \mathbb{R}_{\geq 0}$. Given a bid profile $b = (b_1, \ldots, b_n) \in V$, the auction computes an allocation $g(b)$ and payments $p(b)$. We denote the valuation profile $v$ without bidder $i$ as $v_{-i}$, and similarly for bid profile $b_{-i}$. The set of possible valuation profiles for all bidders except bidder $i$ is denoted as $V_{-i} = \prod_{j \neq i} V_j$.



A bidder's utility, denoted as $u_i(v_i, b)=v_i(g_i(b)) - p_i(b)$, is determined by their valuation $v_i$ and the bid profile $b$. This utility function captures the trade-off between the bidder's valuation for the items and the price they must pay to acquire them. An auction is a dominant strategy incentive compatible (DSIC) if truthful reporting is the optimal strategy for every bidder, regardless of the other bidders' reports.  In formal terms, this means that $u_i(v_i, (v_i, b_{-i})) \geq u_i(v_i, (b_i, b_{-i}))$ holds true for all bidders $i$, all valuations $v_i \in V_i$, all bids $b_i \in V_i$, and all bids $b_{-i} \in V_{-i}$ from others. On the other hand, an auction is said to be individually rational (IR) if every bidder obtains a non-zero utility, meaning that $u_i(v_i, (v_i, b_{-i})) \geq 0$. This ensures that every bidder receives some benefit from participating in the auction and does not suffer a loss. 

A DSIC auction is designed so that each bidder has a dominant strategy to report the true valuation, and the revenue on the valuation profile $v$ is $\sum_i p_i(v)$. 
The goal is to identify an auction mechanism that satisfies DSIC and IR and maximize the revenue.For DSIC   we use the ex-post regret to measure the violation of DSIC, defined as follows:
\begin{eqnarray}\label{Rgt}
Rgt_{i}=\max_{v_{i}^{\prime} \in V_{i}} u_{i}\left(v_{i},\left(v_{i}^{\prime}, v_{-i}\right)\right)-u_{i}\left(v_{i},\left(v_{i}, v_{-i}\right)\right)
\end{eqnarray}

\section{Empirical Analysis of Regret Underestimation in Deep-learning Based Mechanisms}

In this section, we conduct a systematic empirical investigation into the reliability of regret estimation protocols currently employed in deep auction mechanisms. We begin by characterizing the significant heterogeneity in the hyperparameter configurations used by four leading models: RegretNet, ALGnet, RegretFormer, and CITransNet. Subsequently, through rigorous cross-validation and granular sensitivity analysis, we demonstrate that standard evaluation protocols systematically fail to detect significant incentive compatibility violations, particularly in high-dimensional settings. To ensure a fair and reproducible comparison, all experiments utilize the official implementations of the respective models, evaluating their performance initially under their original hyperparameter configurations.

\subsection{Hyperparameter Heterogeneity and Cross-Validation}

We begin by formalizing the workflow of the gradient-based misreport optimizer used to estimate regret. For each valuation profile \( v_i \), the optimizer initializes \( L \) random candidate misreport vectors \( b_i \in V_i \). For each candidate \( b_i^{(l)} \) where \( l \in \{1, \ldots, L\} \), we perform \( R \) steps of projected gradient ascent to maximize the bidder's utility:
\[
b_i^{(l)} \leftarrow  b_i^{(l)} + \gamma \nabla_{b_i} u_i(v_i, (b_i^{(l)}, v_{-i})) ,
\]
where \( \gamma \) is the learning rate. The candidate yielding the highest final utility is selected as the approximate optimal misreport.

\textbf{Lack of Consensus in Evaluation.} 
As shown in Table~\ref{30}, there is no standardized protocol for this evaluation. The hyperparameter configurations employed by these models exhibit substantial variation. For instance, RegretNet uses a computationally intensive setting ($L=1000, R=2000$), whereas RegretFormer uses a minimalist configuration with only a single initialization ($L=1$) and moderate steps, and ALGnet employs a very small learning rate ($\gamma=0.001$).

\begin{table}[t]
  \centering
  
  \begin{tabular}{lrrr}
    \toprule
    \textbf{Model} & $\gamma$  & $L$  & $R$  \\
    \midrule
    RegretNet      & 0.1                      & 1000           & 2000        \\
    ALGnet         & 0.001                    & 300            & 300         \\
    RegretFormer   & 0.1                      & 1              & 1000        \\
    CITransNet     & 0.001                    & 100            & 200         \\
    \bottomrule
  \end{tabular}
  \caption{Hyperparameter configurations across different baseline models}
  \label{30}
\end{table}

\textbf{Cross-Evaluation of Hyperparameters.} 
To investigate whether these model-specific hyperparameters are sufficient to detect incentive compatibility violations, we performed a cross-evaluation experiment in the $3\times10$ setting. We applied the hyperparameter sets from RegretNet, ALGnet, and RegretFormer to each model (CITransNet is excluded here due to its differing valuation distribution). Table~\ref{310} reveals that default hyperparameters systematically underestimate violations. Specifically, RegretNet's intensive settings expose significantly higher regret in ALGnet ($4.78\times 10^{-3}$) than its native configuration ($1.56\times 10^{-3}$). Most critically, RegretFormer exhibits extreme fragility: its detected regret explodes from $5.17\times 10^{-3}$ to $144.02\times 10^{-3}$—a 28-fold increase—when evaluated with ALGnet's finer learning rate. This proves that suboptimal evaluation settings can completely mask catastrophic mechanism failures.

\begin{table}[h]
  \centering
  
  \begin{tabular}{lccc}
    \toprule
    & \multicolumn{3}{c}{\textbf{Hyperparameter Source}} \\
    \cmidrule(lr){2-4}
    \textbf{Model} & RegretNet & ALGnet & RegretFormer\\
    \midrule
    RegretNet    & \textbf{1.62} & 0.40 & 1.55 \\
    ALGnet       & \textbf{4.78} & 1.56 & 3.89 \\
    RegretFormer & 10.59 & \textbf{144.02} & 5.17 \\
    \bottomrule
  \end{tabular}
  \caption{Cross-evaluation of Regret ($\times 10^{-3}$) under different hyperparameter sets ($3\times10$ setting)}
  \label{310}
\end{table}

\subsection{Hyperparameter Sensitivity and Convergence Analysis}
\begin{table*}

  \centering
  \begin{tabular}{rrrrrrrrrr}
    \toprule
    \textbf{Setting}& \textbf{L }    &\textbf{R}& \textbf{Regret ($10^{-3}$)}&\textbf{Run time (h)}&\textbf{Setting}& \textbf{L }    & \textbf{R} &\textbf{Regret($10^{-3}$)}&\textbf{Run time(h)}\\
    \midrule

    \multirow{18}{*}{$3\times10$} &\multirow{6}{*}{$300$}&100&0.14 & 0.6&\multirow{18}{*}{$5\times10$} &\multirow{6}{*}{$300$}&100&0.15 & 3.0 \\
    &&300&1.56 & 2.0&&&300&2.61 & 9.1 \\
    &&500&3.89&3.1&&&500&4.81 & 15.2    \\
    &&1000&4.63&6.3 &&&1000&5.10 & 31.3   \\
    &&1500&4.66&9.6 &&&1500&5.15 & 47.3  \\
    &&2000&4.67&13.9 &&&2000&5.16 & 62.7   \\
    \noalign{\vskip 2pt}
    \cline{2-5}
    
    \cline{7-10}
    \noalign{\vskip 3pt}
    
    &\multirow{6}{*}{$500$}&100&0.17&1.1 &&\multirow{6}{*}{$500$}&100&0.22 & 5.1  \\
    &&300&1.86&3.4&&&300&3.02 & 15.5  \\
    &&500&4.11&5.6 &&&500&4.90 & 26.1  \\
    &&1000&4.67&11.3&&&1000&5.12 & 54.3   \\
    &&1500&4.68&16.2 &&&1500&5.16 &  77.5  \\
    &&2000&4.69&22.1 &&&2000& 5.16 &  104.4  \\
    \noalign{\vskip 2pt}
    \cline{2-5}
    
    \cline{7-10}
    \noalign{\vskip 3pt}
    
    &\multirow{6}{*}{$1000$}&100&0.17&1.1 &&\multirow{6}{*}{$1000$}&100&0.38 & 5.1  \\
    &&300&2.59&6.7&&&300&3.65 & 10.9  \\
    &&500&4.46&11.2 &&&500&5.05 & 54.2  \\
    &&1000&4.77&23.3&&&1000&5.22 & 110.8   \\
    &&1500&4.78&32.5 &&&1500&5.23 &  161.9  \\
    &&2000&4.78&47.6 &&&2000& 5.24 &  216.9  \\

    \bottomrule
  \end{tabular}
    \caption{Compute regret by different hyperparameters on ALGnet}
  \label{300}
\end{table*}

This cross-validation demonstrates that the reported IC properties of existing mechanisms are highly dependent on the specific and often suboptimal choices of the evaluation optimizer. Without a rigorous, unified, and high-strength evaluation standard, the community risks significantly underestimating the extent of incentive compatibility violations.

To further investigate the mechanism of regret underestimation, we conducted a granular sensitivity analysis on the ALGnet optimizer, specifically targeting the number of random initializations ($L$) and gradient descent iterations ($R$). The results, detailed in Table~\ref{300}, track the evolution of computed regret and runtime across varying configurations, revealing a critical trade-off between evaluation fidelity and computational cost that explicitly highlights the inadequacy of default hyperparameters.

\textbf{Sensitivity to Iteration Steps ($R$).} 
With $L$ fixed at 300 (the default for ALGnet), detected regret in the $3\times10$ setting rises monotonically, stabilizing only after $R \approx 1000$. Specifically, increasing $R$ from 100 to 2000 causes a 33-fold jump in regret ($0.14 \times 10^{-3}$ to $4.67 \times 10^{-3}$). Consequently, the default setting of $R=300$ captures merely $\sim 33\%$ of the converged value ($1.56\times 10^{-3}$ vs. $4.67\times 10^{-3}$), creating a misleading illusion of incentive compatibility.

\textbf{Impact of Initialization ($L$).}
Expanding the random pool ($L$) further refines the search. Increasing $L$ from 300 to 1000 pushes the converged regret from $4.67\times 10^{-3}$ to $4.78\times 10^{-3}$. This sensitivity confirms the non-convex nature of the loss landscape, where limited sampling fails to locate the true global maximum.

\textbf{The Prohibitive Cost of Convergence.} 
Achieving these accurate estimates requires immense computational resources. In the $5\times10$ setting, converging to the true regret ($5.24 \times 10^{-3}$) consumes 216.9 hours. In stark contrast, the configuration ($L=300, R=100$) completes in 3 hours but yields a negligible estimate of $0.15 \times 10^{-3}$, underestimating the true violation by a factor of roughly 35.

These experiments conclusively demonstrate that standard gradient-based evaluation is computationally inefficient for large-scale auctions. To obtain reliable regret estimates, one must either invest hundreds of hours or adopt more efficient search strategies like the item-wise guided refinement proposed in this work.

\section{Method}

In this chapter, we present a systematic framework for accurately estimating ex-post regret in deep learning-based auction mechanisms, addressing the critical issue where standard gradient-based evaluators systematically underestimate incentive compatibility violations due to the non-convex loss landscape. We first derive a theoretical lower bound and introduce Item-wise Regret, a computationally efficient proxy that scales linearly with the number of items. Building on this, we propose the Item-wise Guided Gradient Refinement strategy. This hybrid optimization approach utilizes structured initializations to escape local optima effectively and converge to the true global maximum.

\subsection{Mathematical Lower Bound of Regret}

To compute the regret for any bidder as defined in Equation~\ref{Rgt}, we must evaluate all possible valuation profiles $V$. Sampling precision refers to the discretization of the bid space when computing regret. For instance, if the sampling precision is $10^{-3}$ and the bidding range is $[0, 1]$, the interval is divided into 1,000 equally sampling spaced points, each 0.001 apart. Let $Q$ denote the number of sampling points for each bidder and each item in the discretized bid space.

\begin{proposition}\label{pro1}
The time complexity for calculating the optimal regret for all bidders is $O(n \cdot Q^m)$, where $n$ is the number of bidders and $m$ is the number of items.
\end{proposition}
The proof is provided in Appendix A1.

This exponential growth in complexity with respect to the number of items renders the exact computation of optimal regret intractable. To address this, we seek an approximation of Equation~\ref{Rgt} that is both computationally efficient and minimally dependent on hyperparameters.

Instead of using a gradient-based optimizer, we begin with Equation~\ref{Rgt} and decompose the regret across items. We define the regret for bidder $i$ with respect to item $j$ as:

\begin{equation*}
\label{Rgt2}
Rgt_{i,j} = \max_{b_{i,j} \in V_{i,j}} u_i(v_i, (b_{i,j}, v_{i,-j},v_{-i})) - u_i(v_i, (v_i, v_{-i}))
\end{equation*}
where $Rgt_{i,j}$ denotes the regret for bidder $i$ with respect to item $j$, and $v_{i,-j}$ refers to the valuation profile excluding bidder $i$'s valuation for item $j$. This decomposition simplifies the computation while preserving accuracy in regret estimation.

After calculating item-level regret, we define a lower bound of the overall regret for bidder $i$ as:\[Lower\ bound=\max_{j \in M} Rgt_{i,j}.\]

\begin{theorem}\label{the1}
For any bidder \( i \in N \), the following inequality holds:
\[
Rgt_i \geq \max_{j \in M} Rgt_{i,j},
\]
where \( Rgt_i \) is the ex-post regret for bidder \( i \), and \( Rgt_{i,j} \) is the regret of bidder \( i \) on item \( j \).
\end{theorem}

The proof is provided in Appendix A2.

We have therefore introduced a concise and efficient method for computing a lower bound on optimal regret. This method provides a practical way to validate whether regret values produced by gradient-based methods under various hyperparameters $L$ and $R$ are reasonable. If the regret computed via the gradient-based optimizer falls below our theoretical lower bound, it may indicate that the optimizer has failed to identify a valid or representative misreport.

It is worth noting that this lower bound can often closely approximate the true optimal regret. Because many models are trained to reduce regret toward zero, the combinatorial structure of misreports becomes simpler. In such cases, ignoring item-level interactions introduces only minor errors.

\textbf{Tightness and Looseness of the Lower Bound}
The inequality becomes an equality when the optimal misreport of bidder $i$ alters only a single coordinate.  
This situation arises, for example, when the bidder just desires one item, so that only one item can meaningfully increase bidder $i$'s utility.  

If the bidder desires multiple items, the optimal misreport may involve simultaneous deviations on multiple coordinates. 
In this situation, each single-item deviation yields a small gain, making $\text{Lower Bound}$ smaller than $Rgt_i$. Moreover, the discrepancy between $\text{Lower Bound}$ and $Rgt_i$ becomes more pronounced as the number of items that the bidder desires increases.  
In other words, the more items a bidder seeks, the further the lower bound drifts away from the true regret.

\subsection{Item-wise Regret}
To address the issue that the lower bound drifts farther away from the true regret as the number of desired items increases, we introduce the item-wise regret as an alternative approximation, which sums the regret values across all items for each bidder. The definition is:

\begin{equation*}
\label{Rgt3}
\text{Item-wise } Rgt_i =  \sum_{j \in M} Rgt_{i,j}
\end{equation*}

Because the time complexity of computing each $Rgt_{i,j}$ is $Q $, the total complexity of computing item-wise regret for a single bidder is $Q \cdot m$. Therefore, the total complexity across all bidders is $O(n \cdot Q \cdot m)$:

\begin{proposition}\label{pro2}
The time complexity for computing item-wise regret for all bidders is $O(n \cdot Q \cdot m)$, where $n$ is the number of bidders and $m$ is the number of items.
\end{proposition}
The proof is provided in Appendix A3.

This method differs from computing the true optimal regret, as it sums the best misreports for individual items rather than jointly optimizing across all items. Consequently, the item-wise regret may be a little bit larger than the optimal regret. However, it is bounded above and will never exceed $m$ times the optimal regret, where $m$ is the number of items.
\begin{proposition}\label{prop:iw-upper}
If $Rgt_{i,j}\ge 0$ for all $j\in M$, then 
\[
\text{Item-wise } Rgt_i=\sum_{j\in M}Rgt_{i,j}\;\le\; m\cdot \max_{j\in M}Rgt_{i,j}\;\le\; m\cdot Rgt_i.
\]
\end{proposition}

The proof is provided in Appendix A4.

Although the $\text{Item-wise } Rgt_i$ is straightforward to compute, it is not a strict upper bound on $Rgt_i$.  
Since neural network–based auction models are black-box and highly non-convex, pathological cases can arise where $\text{Item-wise } Rgt_i$ substantially underestimates the true regret.  
Such extreme deviations occur only when the learned mechanism severely violates incentive compatibility, leading to strong cross-item interactions.  
In practice, however, when cross-item interactions are weak, $\text{Item-wise } Rgt_i$ is typically close to $Rgt_i$ and thus provides a useful empirical proxy.  
Moreover, because most learned mechanisms remain approximately incentive-compatible, our method usually produces estimates that are very close to the optimal regret.  

\subsection{Item-wise Guided Gradient Refinement}

The item-wise regret provides a computationally efficient method. But it relies on a discretized bid space, which inherently suffers from two limitations: (1) the \textit{discretization gap}, where the true continuous optimum lies between grid points, and (2) the \textit{independence assumption}, which may overlook complex multi-item synergies. Conversely, standard gradient-based methods operate in continuous space but requiring massive random restarts (e.g., $N \geq 1000$) to converge to near true regret.

To bridge this gap, we propose a hybrid optimization strategy that utilizes the item-wise results to construct a structured, high-quality initialization portfolio for the gradient-based optimizer. We construct an initialization set $\mathcal{B}_{init}$ of size $K =1+ m + 3*k$ for each bidder $i$. Let $v_i$ denote the truthful valuation and $\hat{b}_{i} = (\hat{b}_{i,1}, \dots, \hat{b}_{i,m})$ denote the vector of optimal item-wise misreports identified by the discrete grid search. The set $\mathcal{B}_{init}$ is composed of the following five strategic components:

\begin{enumerate}
\item \textbf{Combinatorial Candidate (1 candidate):} 
We construct the candidate $\hat{b}_{i}$ by directly aggregating the optimal misreports identified for each individual item. This vector serves as a strong heuristic for the joint strategy, specifically designed to capture the multi-item synergy within the valuation landscape.
    
\item \textbf{Single-Item Candidates ($m$ candidates):} 
These candidates are designed based on the item-wise regret identified during the grid search. Specifically, for each item $j$, we construct a dedicated initialization vector that adopts the optimal misreport $\hat{b}_{i,j}$ for that specific item, while keeping all other dimensions truthful:
\[
b^{(j)} = (\hat{b}_{i,j}, v_{i,-j}), \quad \forall j \in \{1, \dots, m\}.
\]
This design ensures that the specific contribution of each item's locally optimal strategy is explicitly incorporated into the initialization set, directly corresponding to the calculation of Item-wise Regret.
    
\item \textbf{Perturbed Combinatorial Candidates ($k$ candidates):} 
To address potential limitations arising from the independent calculation of item-wise regrets, we generate candidates by applying Gaussian perturbations to the Combinatorial Candidate ($\hat{b}_{i}$). We inject noise into this joint vector to explore its local neighborhood:
\[
b^{(refine)} = \hat{b}_{i} + \epsilon, \quad \text{where } \epsilon \sim \mathcal{N}(0, \sigma_{opt}^2 I).
\]
    
\item \textbf{Perturbed Truthful Candidates (k candidates):} 
Learned mechanisms often exhibit instability specifically near the truthful valuation. We include candidates generated by perturbing the truthful profile:
\[
b^{(local)} = v_i + \epsilon, \quad \text{where } \epsilon \sim \mathcal{N}(0, \sigma_{truth}^2 I).
\]
    
\item \textbf{Global Random Candidates (k candidates):} 
To ensure robustness against pathological loss landscapes where heuristic priors might fail, we retain a small number of uniformly random initializations $b^{(rand)} \sim U[0, 1]^m$ as a safety mechanism.
    
\end{enumerate}

This structured initialization drastically lowers the sample requirements. In experiments, it only needs a small number of initializations rather than necessitating thousands of random initializations in standard approaches.

\section{Experiments}

In this section, we present a comprehensive empirical evaluation of our proposed Item-wise Guided Gradient Refinement framework. Our experiments are designed to verify two core claims: accuracy in approximating the true optimal regret, efficiency in reducing computational overhead. 
We validate our approach on the ALGnet benchmark, demonstrating that our method achieves fidelity comparable to the near-optimal regret while delivering orders-of-magnitude speedups over the baseline. 
We then apply our framework to reevaluate three leading deep auction models: RegretNet, ALGnet, RegretFormer and CITransNet. Our findings reveal a spectrum of reliability: while RegretNet remains robust, we detect moderate regret underestimation in ALGnet and CITransNet and expose catastrophic hidden violations in RegretFormer. 

To ensure a fair comparison, we maintained the original hyperparameter configurations for all models and focused on re-evaluation. We report the average results over 10,000 test samples. On RegretNet, ALGnet and RegretFormer, We examine \( n \)-bidder \( m \)-item additive settings, where the valuations are sampled i.i.d from \( U[0, 1] \), denoted as \( n \times m \). As for CITransNet, its valuation is more complex; we show the result in Appendix B. The evaluation on ALGnet shows that our method can be used for all distributions of bidders' valuations. The sampling precision is set to \(10^{-3}\), meaning each bidder's bid space for each item is uniformly divided into \(10^3\) sampling values. Based on previous research\cite{you2023improving}, increasing the sampling precision beyond this level does not yield higher regret values but significantly increases computation time.

\subsection{Hyperparameter Selection and Rationale}
In this section, we detail the hyperparameter selection for our guided refinement strategy, as summarized in Table~\ref{tab:hyper_noise}. Specifically, we define the Gaussian noise scale $\sigma$ and the group size $k$ assigned to each of the three randomized initialization sets: Perturbed Combinatorial, Perturbed Truthful, and Global Random Candidates. 

\begin{table}[t]
  \centering
  \caption{Hyperparameter settings of $k$ and Gaussian noise ($\sigma$) for different models}
  \label{tab:hyper_noise}
  \begin{tabular}{lcc}
    \toprule
    \textbf{Model} & \textbf{$k$} & \textbf{Gaussian Noise ($\sigma$)} \\
    \midrule
    RegretNet    & 0  & 0   \\
    ALGnet       & 0  & 0   \\
    RegretFormer & 80 & 0.6 \\
    \bottomrule
  \end{tabular}
\end{table}

Notably, for RegretNet and ALGnet, the optimizer achieves robust convergence relying solely on the deterministic priors—namely, the single Combinatorial Candidate and the $m$ Single-Item Candidates (where $m$ denotes the number of items). Consequently, we set $k=0$ for these models. In sharp contrast, RegretFormer necessitates a larger candidate pool ($k=80$) to converge. This disparity arises because RegretFormer severely violates incentive compatibility, inducing a highly complex misreport landscape. Therefore, a diverse set of randomized initializations is indispensable to preventing the optimizer from being trapped in local optima and ensuring the detection of near true regret.

\begin{table}[t]
\centering

    \begin{tabular}{llrr} 
            \toprule
              \textbf{Setting}& \textbf{Metric}  & \textbf{Regret $(\times 10^{-3})$}&\textbf{Time (h)}\\
              \midrule
            
            \multirow{4}{*}{$1\times2$}
            &\makecell[l]{Original} &0.27&0.11 \tabularnewline
            &\makecell[l]{Optimal} &0.28&10.5 \tabularnewline 
            &\makecell[l]{Lower Bound} &0.23 &0.003\tabularnewline 
            &\makecell[l]{Item-wise} &0.36&0.003 \tabularnewline 
            &\makecell[l]{\textbf{Ours}} &\textbf{0.28}&\textbf{0.012} \tabularnewline 
            
             \midrule
             
              \multirow{4}{*}{$2\times2$}
              &\makecell[l]{Original} &0.71&0.85 \tabularnewline 
              &\makecell[l]{Optimal} &0.72&25.5 \tabularnewline
            &\makecell[l]{Lower Bound} &0.59 &0.005\tabularnewline 
            &\makecell[l]{Item-wise} &0.77&0.005 \tabularnewline 
            &\makecell[l]{\textbf{Ours}} &\textbf{0.71}&\textbf{0.025} \tabularnewline 

             \midrule
             
             \multirow{4}{*}{$5\times2$}
             &\makecell[l]{Original} &0.38 &4.9\tabularnewline 
             &\makecell[l]{Optimal} &0.39&86.0 \tabularnewline 
            &\makecell[l]{Lower Bound} &0.32 &0.03\tabularnewline 
            &\makecell[l]{Item-wise} &0.39&0.03 \tabularnewline 
            &\makecell[l]{\textbf{Ours}} &\textbf{0.38}&\textbf{0.15} \tabularnewline 
            
            \midrule
            
            \multirow{4}{*}{$3\times10$}
            &\makecell[l]{Original} &1.56&2.0 \tabularnewline 
            &\makecell[l]{Near Optimal} &4.72&4.26 \tabularnewline             
            &\makecell[l]{Lower Bound} &2.17 &0.09\tabularnewline 
            &\makecell[l]{Item-wise} &6.28&0.09 \tabularnewline 
            &\makecell[l]{\textbf{Ours}} &\textbf{4.73}&\textbf{0.27} \tabularnewline 

            \midrule
            
            \multirow{4}{*}{$5\times10$}
            &\makecell[l]{Original} &2.48 &9.2\tabularnewline
            &\makecell[l]{Near Optimal} &5.24&216.9 \tabularnewline
            &\makecell[l]{Lower Bound} &2.49&0.57 \tabularnewline 
            &\makecell[l]{Item-wise} &5.50&0.57 \tabularnewline
            &\makecell[l]{\textbf{Ours}} &\textbf{5.25}&\textbf{1.19} \tabularnewline 
            
             \bottomrule
    \end{tabular}
      \caption{Comparison of Regret Metrics and Runtime on ALGnet}\label{result2}
\end{table}

\begin{table*}[t]
\centering

	\begin{tabular}{cccrrrrr} 
            \toprule
              \textbf{Setting}& \textbf{Model}& $R_{max}$ & \textbf{ \makecell{Regret $(\times 10^{-3})$ \\ (original)} }&\textbf{ \makecell{Regret $(\times 10^{-3})$ \\ (Ours)} }&\textbf{Revenue}\\
              \midrule
\multirow{4}{*}{$1\times2$}
            &\makecell[l]{RegretNet} &-&0.23&\textbf{0.23}&\textbf{0.553} \tabularnewline 
             &\makecell[l]{ALGnet}&- &0.27&\textbf{0.28}&\textbf{0.545} \tabularnewline
             &\multirow{2}{*}{RegretFormer}&$1\times10^{-3}$ &1.21&\textbf{1.71}&\textbf{0.575} \tabularnewline
             &&$1\times10^{-4}$&0.09&\textbf{0.21}&\textbf{0.555} 
             \tabularnewline 
             \midrule
              \multirow{4}{*}{$2\times2$}
            &\makecell[l]{RegretNet}&- &0.26 &\textbf{0.26}&\textbf{0.878 }\tabularnewline 
            &\makecell[l]{ALGnet}&- &0.71&\textbf{0.72}&\textbf{0.880} \tabularnewline
              &\multirow{2}{*}{RegretFormer}&$1\times10^{-3}$ &0.59&\textbf{21.65}&\textbf{0.902} \tabularnewline
             &&$1\times10^{-4}$ &0.04&\textbf{18.31}&\textbf{0.862} \\
             \midrule

             \makecell[c]{$2\times5$}&\makecell[l]{RegretFormer} &{$1\times10^{-3}$}&1.54&\textbf{42.55}&\textbf{2.46} \tabularnewline  
            \midrule
              \multirow{3}{*}{$3\times10$}
            &\makecell[l]{RegretNet} &-&1.62 &\textbf{1.62}&\textbf{5.54} \tabularnewline
            &\makecell[l]{ALGnet}&- &1.56&\textbf{4.73}&\textbf{5.55 }\tabularnewline
              &\makecell[l]{RegretFormer}&$1\times10^{-3}$ &5.17&\textbf{365.80}&\textbf{6.18} \tabularnewline
             \midrule
            
            \multirow{2}{*}{$5\times10$}
            &\makecell[l]{RegretNet}&- &4.72 &\textbf{5.25}&\textbf{6.76} \tabularnewline 
            &\makecell[l]{ALGnet}&- &2.48 &\textbf{5.19}&\textbf{6.74 }\tabularnewline 
            \bottomrule
    \end{tabular}
    \caption{Evaluate regret on RegretNet, ALGnet and RegretFormer}\label{resultall}
 \end{table*}

\subsection{Validating Our Approach on ALGnet}
To assess the effectiveness and efficiency of our proposed framework, we conducted a comprehensive evaluation on the ALGnet model. The results are summarized in Table~\ref{result2}. We compare five distinct metrics: (1) Original: The regret calculated by the model's original parameters. (2) Optimal / Near Optimal: The ground truth calculated via exhaustive search in small settings or calculated by converging a gradient-based optimizer with intensive settings (1,000 random initializations and 2,000 iterations) in large settings. (3 )Lower Bound: Proposed in Section 4.1. (4) Item-wise: Proposed in Section 4.2. (5) Ours: The Item-wise Guided Gradient Refinement method proposed in Section 4.3. 

In small-scale settings ($1\times2$, $2\times2$, and $5\times2$) where the true optimal regret is computable via exhaustive search, our framework demonstrates both exceptional precision and efficiency. The \textit{Lower Bound} and \textit{Item-wise} regret consistently bracket the true \textit{Optimal} regret; for instance, in the $2\times2$ setting, the optimal value ($0.72\times 10^{-3}$) is tightly sandwiched between the lower bound ($0.59\times 10^{-3}$) and the item-wise proxy ($0.77\times 10^{-3}$). And our proposed method (\textbf{Ours}) achieves estimates almost identical to the ground truth (e.g., $0.71\times 10^{-3}$ vs. $0.72\times 10^{-3}$ in $2\times2$). Moreover, this high fidelity is achieved with negligible computational cost: in the $5\times2$ case, our method requires only 0.15 hours compared to the 86 hours needed for exhaustive search, achieving a speedup of nearly three orders of magnitude.

In high-dimensional settings ($3\times10$, $5\times10$) where true optima are intractable, we utilize a "Near Optimal" benchmark derived from intensive gradient search (1,000 initializations, 2,000 iterations). Against this standard, our method demonstrates superior robustness and efficiency. First, we expose the failure of standard baselines: in the $3\times10$ case, the original optimizer reports a regret ($1.56\times 10^{-3}$) paradoxically lower than our theoretical lower bound ($2.17\times 10^{-3}$), mathematically proving that it misses even basic single-item violations. Second, regarding efficiency, obtaining the benchmark result in $5\times10$ requires 216.9 hours. In contrast, our method achieves a comparable estimate ($5.25\times 10^{-3}$) in just 1.19 hours. By leveraging item-wise priors, our framework achieves a $\sim 200\times$ speedup, transforming rigorous verification from a week-long task into one feasible within hours.

\subsection{Re-evaluating State-of-the-Art Models}

We applied our proposed evaluation framework to re-examine the incentive compatibility of three leading deep auction mechanisms: RegretNet, ALGnet, and RegretFormer. Table~\ref{resultall} compares the regret reported under their original hyperparameter configurations against the values detected by our guided refinement strategy.

The results reveal a significant dichotomy in model robustness. 
RegretNet demonstrates high stability, with our method yielding regret estimates nearly identical to its original reports across all settings (e.g., $1.62\times 10^{-3}$ vs. $1.62\times 10^{-3}$ in $3\times10$), confirming that its intensive evaluation protocol effectively captures true violations. 
ALGnet, however, exhibits moderate underestimation in high-dimensional settings; for instance, in the $3\times10$ case, our method detects a regret of $4.73\times 10^{-3}$, approximately three times higher than the reported $1.56\times 10^{-3}$.

Most critically, our evaluation uncovers catastrophic hidden violations in RegretFormer. While the model claims high revenue and low regret (e.g., $5.17\times 10^{-3}$ in $3\times10$) under its default settings, our method exposes a massive regret of $365.80\times 10^{-3}$—an increase of over 40 times. This pattern is consistent across settings (e.g., $1.54\times 10^{-3}$ vs. $42.55\times 10^{-3}$ in $2\times5$), indicating that RegretFormer's high reported revenue is likely achieved by sacrificing incentive compatibility, a failure completely masked by its original, insufficient evaluation protocol. These findings underscore the necessity of our high-strength evaluation framework for ensuring the reliability of learned mechanisms.

\section{Conclusion}

In this paper, we first conduct a systematic evaluation of existing deep learning–based auction models, including RegretNet, ALGnet, RegretFormer, and CITransNet. Our experiments reveal that the regret reported by several models is often lower than the actual regret, particularly in multi-item settings. This underestimation suggests that part of the reported revenue may stem from undetected violations of incentive compatibility. Motivated by these findings. To address this issue, we derive a lower bound on regret and introduce an efficient item-wise regret approximation. Building on this, we propose a guided refinement procedure that substantially improves regret estimation accuracy while reducing computational cost.  These methods enable more accurate regret assessment at substantially lower computational cost. Overall, our work both exposes key limitations of current practice and provides principled and practical tools for evaluating incentive properties in deep learning–based auction mechanisms.

\bibliographystyle{named}
\bibliography{ijcai26}

\newpage
\section*{Technical Appendix}

\subsection{Proof of Proposition 1}

\begin{proposition}\label{pro1}
The time complexity for calculating the optimal regret for all bidders is $O(n \cdot Q^m)$, where $n$ is the number of bidders and $m$ is the number of items.
\end{proposition}

\begin{proof}
For each bidder $i$, computing the regret requires considering all possible alternative bids across all $m$ items while keeping the other bidders' bids fixed.  
Since we discretize each item's valuation into $Q$ equally spaced points, the number of possible combinations of valuations across $m$ items is $Q \times Q \times \cdots \times Q = Q^m$.  

Therefore, for a single bidder, $Q^m$ evaluations are needed. Considering all $n$ bidders, the total number of evaluations becomes $n \cdot Q^m$. Hence, the time complexity for calculating the optimal regret for all bidders is $O(n \cdot Q^m)$.
\end{proof}

\subsection{Proof of Theorem 1}
\begin{theorem}\label{the1}
For any bidder \( i \in N \), the following inequality holds:
\[
Rgt_i \geq \max_{j \in M} Rgt_{i,j},
\]
where \( Rgt_i \) is the ex-post regret for bidder \( i \), and \( Rgt_{i,j} \) is the regret of bidder \( i \) on item \( j \).
\end{theorem}
\begin{proof}
By definition, the overall ex-post regret of bidder \( i \) is:
\[
Rgt_i =  \max_{b_i \in V_i} u_i(v_i, (b_i, v_{-i})) - u_i(v_i, (v_i, v_{-i})).
\]

We decompose the full misreport \( b_i \) into two parts: \( b_{i,j} \) for item \( j \), and \( b_{i,-j} \) for all items other than \( j \). Hence, \( b_i = (b_{i,j}, b_{i,-j}) \). Then  

\begin{equation}
\small 
    Rgt_i = \max_{(b_{i,j}, b_{i,-j}) \in V_i}( u_i(v_i, ((b_{i,j}, b_{i,-j}), v_{-i})) - u_i(v_i, (v_i, v_{-i})) ).
\end{equation}

Let \( j^* = \arg\max_{j \in M} Rgt_{i,j} \), and let \( b_{i,j^*}' \in V_{i,j^*} \) denote the optimal misreport for item \( j^* \), that is:
\[
b_{i,j^*}' = \arg\max_{b_{i,j} \in V_{i,j}} ( u_i(v_i, (b_{i,j}, v_{i,-j})v_{-i}) - u_i(v_i, (v_i, v_{-i})) ).
\]
where $v_{i,-j}$ refers to the valuation profile excluding bidder $i$'s valuation for item $j$

We construct a restricted misreport vector where the value for item \( j^* \) is set to \( b_{i,j^*}' \), and all other items are kept truthful. Therefore, we have:
\begin{equation}
\small
Rgt_i \geq  \max_{b_{i,-j^*} \in V_{i,-j^*}} u_i(v_i, ((b_{i,j^*}', b_{i,-j^*}), v_{-i})) - u_i(v_i, (v_i, v_{-i})) 
\end{equation}

The inequality holds because we restrict the optimization domain. And we have:
\[
Rgt_i \geq  u_i(v_i, ((b_{i,j^*}', v_{i,-j^*}), v_{-i})) - u_i(v_i, (v_i, v_{-i})).
\]

The inequality holds because we also take a specific (not necessarily optimal) value for \( b_{i,-j^*} \), namely the truthful value \( v_{i,-j^*} \).

Since only \( b_{ij^*}' \) is misreported, and all other values remain truthful. Thus:
\[
\begin{aligned}
Rgt_i &\geq  u_i(v_i, (b_{ij^*}', v_{-ij^*})) - u_i(v_i, (v_i, v_{-i}))  \\
& = \max_{j \in M} Rgt_{i,j} =Lower\ bound.
\end{aligned}
\]

This completes the proof.
\end{proof}

\subsection{Proof of Proposition 2}
\begin{proposition}\label{pro2}
The time complexity for computing item-wise regret for all bidders is $O(n \cdot Q \cdot m)$, where $n$ is the number of bidders and $m$ is the number of items.
\end{proposition}

\begin{proof}
For each bidder $i$, the item-wise regret is defined as the sum of the regrets for all $m$ items, i.e., $\text{Item-wise } Rgt_i = \sum_{j \in M}^M Rgt_{i,j}$ as in Equation~\eqref{Rgt3}.  

Computing $Rgt_{i,j}$ for a single item $j$ requires evaluating $Q$ possible bid values for that item. Since there are $m$ items, computing the item-wise regret for a single bidder requires $Q \cdot m$ evaluations.  

Considering all $n$ bidders, the total number of evaluations becomes $n \cdot Q \cdot m$. Therefore, the time complexity for computing item-wise regret for all bidders is $O(n \cdot Q \cdot m)$.
\end{proof}

\subsection{Proof of Proposition 3}
\begin{proposition}\label{prop:iw-upper}
If $Rgt_{i,j}\ge 0$ for all $j\in M$, then 
\[
\text{Item-wise } Rgt_i=\sum_{j\in M}Rgt_{i,j}\;\le\; m\cdot \max_{j\in M}Rgt_{i,j}\;\le\; m\cdot Rgt_i.
\]
\end{proposition}
\begin{proof}
For any bidder $i$ and item $j\in M$, by Theorem \ref{the1}, we can find that: the deviation space of misreporting only item $j$ is a subset of the full deviation space that defines $Rgt_i$. Hence,
\[
\max_{j\in M} Rgt_{i,j} \le Rgt_i.
\]

By the inequality $\sum_{j=1}^m a_j \le m \cdot \max_j a_j$ for nonnegative numbers $a_j$, we obtain
\[
\text{Item-wise } Rgt_i = \sum_{j=1}^m Rgt_{i,j} \le m \cdot \max_{j\in M} Rgt_{i,j} \le m \cdot Rgt_i.
\]

Therefore, the item-wise regret is at most $m$ times the optimal regret. Equality can occur in extreme cases where each single-item misreport yields the same regret as the joint misreport, but in general, the bound is strict.
\end{proof}

\section{Results on CITransNet}
Since the CITransNet model incorporates additional contextual information, the valuation distribution of bidders over items differs from that used in the previous three models. Therefore, we present its results separately in this section.

To ensure fairness, we use two complex settings for testing:

(D) 2 bidders with $X = \{1,2,\dots,10\}$ and 5 items with $Y = \{1,2,\dots,10\}$. All the contexts are uniform sampled, and $v_{ij}$ is drawn according to the normal distribution $v_{ij} \sim \mathcal{N}\Bigg(\frac{(x_i+y_j)\bmod 10 + 1}{11}, 0.05\Bigg) \text{truncated to }  [0,1].$

(E) 3 bidders and 10 items. The discrete contexts and corresponding values are drawn similarly as Setting D.
\begin{table}
\centering
  \caption{Test regret on CITransNet}\label{CITransNet}

	\begin{tabular}{llr} 
            \toprule
              \textbf{Setting}& \textbf{Method}  & \textbf{ Regret $(\times 10^{-3})$}\\
              \midrule
\multirow{2}{*}{(D)}
            &\makecell[l]{CITransNet} &0.11 \tabularnewline 
            &\makecell[l]{Ours} &0.69 \tabularnewline 
             \midrule
              \multirow{2}{*}{(E)}
            &\makecell[l]{CITransNet} &0.24 \tabularnewline 
            &\makecell[l]{Ours} &4.82 \tabularnewline 
             
            \bottomrule
    \end{tabular}
\end{table}

Table~\ref{CITransNet} reports the test regret of CITransNet under two complex auction settings. We observe that the regret reported by CITransNet is significantly lower than the regret calculated by our method. For example, in setting (D), CITransNet yields a regret of 0.11, whereas our regret is 0.69; in setting (E), the reported regret is 0.24, whereas our result is 4.82. 

This consistent underestimation indicates that the misreport optimization process in CITransNet does not fully converge. The main cause lies in its hyperparameter configuration—specifically, the number of gradient steps and the number of initialization misreport sets are set too small. As a result, the optimization terminates prematurely, failing to reach the local or global optimum of the regret landscape. This issue becomes increasingly severe as the number of bidders and items grows, since the misreport search space expands exponentially and the optimization becomes more challenging.

In summary, while CITransNet incorporates contextual information and demonstrates strong representational capability, its reported regret significantly underestimates true IC violations due to inadequate misreport optimization. This underestimation is exacerbated by the problem scale, as convergence becomes prohibitively slow. In contrast, our method effectively addresses these limitations and accurately evaluates the true regret.

\end{document}